\documentclass{article}

\usepackage{arxiv}

\usepackage[utf8]{inputenc} 
\usepackage[T1]{fontenc}    
\usepackage{hyperref}       
\usepackage{url}            
\usepackage{booktabs}       
\usepackage{amsfonts}       
\usepackage{nicefrac}       
\usepackage{microtype}      
\usepackage{lipsum}
\usepackage{graphicx}
\usepackage{dcolumn}
\usepackage{bm}

\title{Multiphoton absorption excited upconversion luminescence in optical fiber}

\author{\\F. Mangini$^1*$, M. Ferraro$^2*$, M. Zitelli$^2$, A. Niang$^1$, A. Tonello$^3$, V. Couderc$^3$, S. Wabnitz$^{2,4}$}

\begin{document}
\maketitle

$^1$ Department of Information Engineering (DII), University of Brescia, Via Branze 38, 25123
 Brescia, Italy.\\
$^2$ Department of Information Engineering, Electronics and Telecommunications (DIET),
Sapienza University of Rome, Via Eudossiana 18, 00184 Rome, Italy.\\
$^3$ Universit\'e de Limoges, XLIM, UMR CNRS 7252, 123 Avenue A. Thomas, 87060 Limoges, France\\
$^5$  Novosibirsk State University, Pirogova 1, Novosibirsk 630090, Russia\\
$^*$ These authors have contributed equally

\maketitle

\begin{abstract}
We experimentally demonstrate a novel nonlinear effect in optical fibers: upconversion luminescence generation excited by multiphoton absorption of femtosecond infrared pulses. We directly estimate the average number of photons involved in the up-conversion process, by varying the wavelength of the pump source. We highlight the role of non-bridging oxygen hole centers and oxygen deficient center defects, and directly compare the intensity of side-scattered luminescence with numerical simulations of pulse propagation.
\end{abstract}

\section*{}
Upconversion  luminescence  (UL),  involving  the  absorption  of  multiple  photons  followed  by  the  emission of higher energy photons via the radiative decay of excited  electron  energy  levels  in  dielectric  materials,  has attracted  considerable  research  attention,  since  its  first observation in rare-earth-doped crystals \cite{varsanyl}. Over the years, UL found application in a wide range of fields, moving from physics to chemistry and biology, since it allows for the detection of several stimuli such as temperature, electromagnetic radiation and pH \cite{tsang}. Thanks to materials nanoengineering development, UL has been recently proposed for medical therapy too \cite{yao}. In physics, UL finds its main application in the generation of visible laser emission by means of infrared optical pumping \cite{pollak,johnson}.
Aiming to the realization of fiber lasers, various authors have studied UL in optical fibers, doped with different types of semiconductors, such as cesium, europium, and tantalum \cite{you,you2,meng}. In their first report of UL in silica glass, which is the main constituent of commercial optical fibers, Kazansky et al. attributed their observation of multiphoton absorption (MPA) excited upconversion luminescence in Ge-doped silica glass to the presence of Germanium Oxygen deficient center (Ge-ODC) defects \cite{kazansky}. However, intrinsic and induced defects in silica were already studied by using ultraviolet (UV)-pumped visible photoluminescence. Among these, the presence of the so-called non-bridging Oxygen hole center (NBOHC) defects, responsible for visible red light emission at 650 nm, attracted significant interest \cite{skuja-1994}. NBOHCs consist of a complementary pair of broken silicon-oxygen bonds, and were among the first defects to be studied. Their paramagnetic properties permit to investigate NBOHCs by means of the electron paramagnetic resonance technique. Only later, the presence of diamagnetic defects, such as Ge-OCDCs, was also discovered in silica glass \cite{skuja-1998}.
These defects are responsible for several luminescence bands in the UV-VIS spectral range. Three main emission bands, centered at 290, 400 and 460 nm, were easily detected in as-grown or irradiated silica \cite{cannas, messina-2011}, as well as in Ge-doped silica \cite{kristensen}. Ge-ODC luminescence was also observed in singlemode optical fibers \cite{gallagher}, and it was proposed as an alignment marker for the manufacturing of fiber Bragg gratings \cite{hnatovsky}. Moreover, defects and impurities luminescence represents a useful tool for the spectroscopy of transparent media, and permits to extend the range of laser emission into the UV. Luminescence was recently proposed as a tool for supercontinuum generation in YAG crystals \cite{kudarauskas}. In optical fibers, UL provides an interesting new means to extend the supercontinuum bandwidth into the UV, which is otherwise challenging by using parametric nonlinear effects \cite{krupa, wright-prl}. 
In this Letter, we report the experimental observation and theoretical description of visible UL in commercial multimode fibers (MMFs), pumped by intense femtosecond infrared laser pulses. Owing to their large numerical aperture, MMFs are better suited than singlemode fibers to trap luminescent radiation. Moreover, because of their relatively large core size, MMFs may carry light beams with peak powers up to the self-focusing critical value, before any permanent damage occurs. This allows for the observation of a host of novel nonlinear effects \cite{krupa, wright-natph}. Specifically, the presence of photoluminescence and nonlinear losses, which were ascribed to a MPA mechanism, was recently reported in graded-index (GRIN) MMFs \cite{zitelli, hansson}. These nonlinear losses introduce a fundamental, and previously undisclosed, nonlinear limitation to the energy transmission capabilities of optical fibers, and may provide an intrinsic limitation for the power scaling of spatiotemporal mode-locking with multimode fiber lasers \cite{wright-science}.
To date, a microscopic interpretation of these intriguing phenomena is still missing. Here we fill this gap of knowledge, by associating them to UL. Specifically, we unveal a new, previously unforeseen nonlinear phenomenon in optical fibers, by demonstrating the presence of up to 5-photon absorption processes involved in UL generation. Spatial self-imaging (SSI) in GRIN fibers plays key role in enhancing UL, which becomes visible to the naked eye as an array of side-scattering blue emitters. Our experimental results are in excellent quantitative agreement with numerical simulations, based on a generalized nonlinear Schr\"{o}dinger equation (GNLS). We provide a complete characterization of the multiphoton nature of UL, by varying the laser power and wavelength, and compare UL spectra for GRIN and step-index MMFs.

\begin{figure}[htbp]
  \centering
  \includegraphics[width=8.6cm]{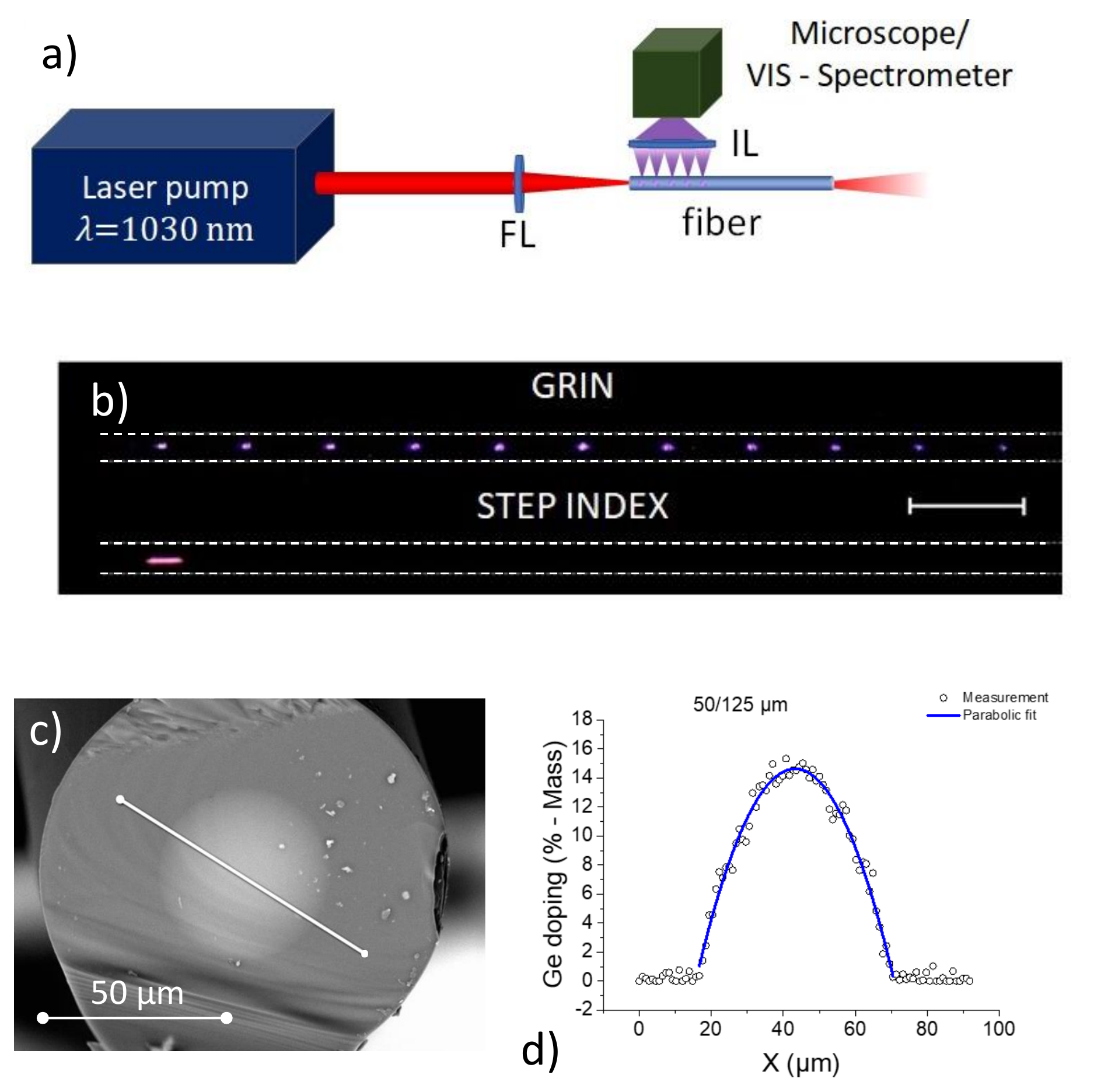}
\caption{(a) Schematic of the experimental setup. (b) Microscope image of luminescence scattered from either GRIN or step-index fibers. Dashed lines stand for the cladding-air interface, while the scale bar corresponds to 1 mm. (c) SEM image of a GRIN fiber section. (d) \% mass Germanium concentration in the Ge-doped GRIN fiber.}
  \label{setup}
\end{figure}

In our experiments, we measured the spectrum of visible UL light scattered out of the first few millimeters of propagation in MMFs. This light results from the multiphoton absorption of high peak power (up to 3 MW) femtosecond infrared pulses. Spectra were collected with a miniature fiber optics spectrometer (Avaspec-2048). As a pump source, we used a hybrid optical parametric amplifier (OPA) pumped by a femtosecond Yb-based laser. Pulses of 80 fs at 650-940 nm and 180 fs at 1030 nm with 100 kHz repetition rate were coupled into two different multimode standard fibers: a 50/125 GRIN fiber with a relative index difference $\Delta_{GRIN}= 0.0103$, and a 105/125 step-index fiber with $\Delta_{SI}=0.012$. In both cases, the beam ($1/e^2$) diameter at the fiber input facet was equal to 7.55 $\mu$m at $\lambda = 1030$ nm, and about 11$\mu$m at $\lambda = 650-940$ nm. UL was collected by means of an imaging lens into either the spectrometer or a microscope. In Fig.\ref{setup}a we present a sketch of our setup. Fig.\ref{setup}b shows pictures of the side-scattered UL spots, taken by a Dinolite-AM3113T digital microscope. The different scattering patterns from either GRIN or step-index fibers can be clearly appreciated. A single emission spot at the point of minimum beam waist is scattered from the step-index fiber. Whereas the periodic beam focusing in the GRIN fiber leads to an array of scattering points. The two fibers also differ in terms of their dopants and concentrations. Specifically, in the step-index fiber the cladding is made of fluorine-doped silica, while the core is nominally undoped. In contrast, index grading in the GRIN fiber is obtained by using a parabolically decreasing concentration of Ge dopant in the core, and the cladding is made of pure silica. In Fig.\ref{setup}c, a scanning electron microscope (SEM) image of the GRIN fiber facet is reported. We spatially tracked the Ge concentration along a diameter of the core (white line in the figure) by means of energy dispersive X-ray Spectroscopy (EDX), which provides a well-known signature in the electromagnetic spectrum. The result is shown in Fig.\ref{setup}d. At its peak, Ge-doping is 15 \%-mass (corresponding to 4.5 \% -atomic). 

\begin{figure}[htbp]
  \centering
  \includegraphics[width=8.6cm]{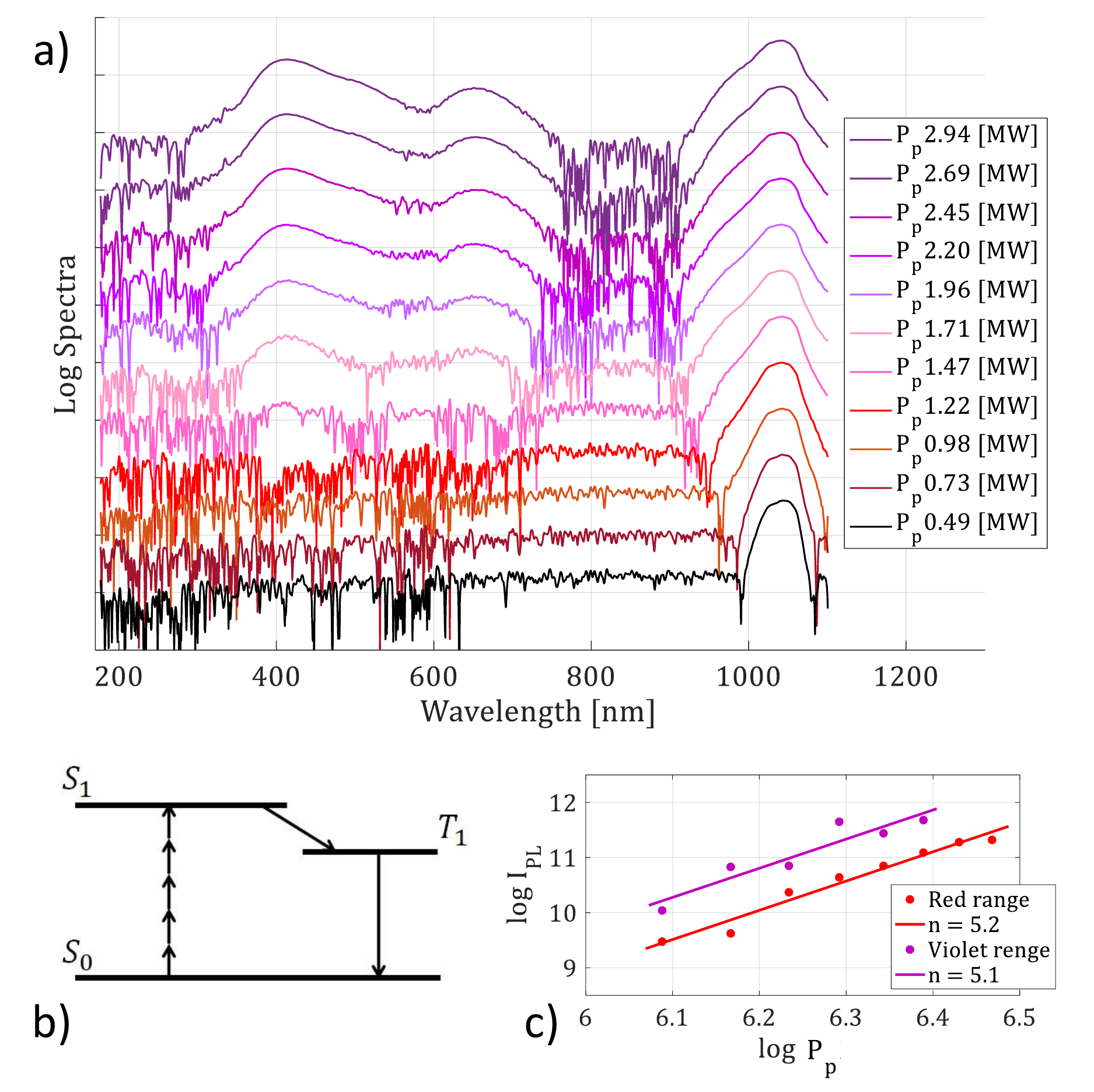}
\caption{(a) UL spectra in log scale when varying the input peak power. (b) Sketch of a five-photon absorption UL mechanism. (c) Log-log plot of UL intensity as a function of input beam peak power. Experimental data (circles) are fitted by solid lines, whose slope $n$ is written in the legend.}
  \label{Spect_mod_n}
\end{figure}

In Fig.\ref{Spect_mod_n}a we illustrate the dependence of the measured spectrum of side-scattered light on the peak power ($P_p$) of the pump laser pulses at 1030 nm. As can be seen, a broad visible emission appears when $P_p$ crosses a certain power threshold (around 1.5 MW). Beyond such power, one may distinguish the generation of three main spectral peaks at 650, 460, and 400 nm, respectively. Another peak at 345 nm is appreciable at the highest input powers. The latter, having a rather narrow bandwidth, can be easily identified with the third-harmonic of the pump laser. On the other hand, we may attribute the origin of the red peak at 650 nm to the presence of NBOHCs. Whereas the peaks at 460 and 400 nm can be ascribed to Ge-ODC. All of these peaks resuls from MPA of the pump source. We did not observe any luminescence peak at 240 nm, since it is negligible at room temperature \cite{leone}. A sketch of the MPA process involving $n = 5$ photons is shown in Fig.\ref{Spect_mod_n}b. Owing to MPA, an electron is excited into a higher energy band ($S_0 \rightarrow S_1$). Next, electron-phonon scattering allows the system to relax into an intermediate energy level, from which luminescence takes place, thus bringing back the electron to its ground state ($T_1 \rightarrow S_0$). The only constraint in this simple picture is that the number of photons involved must be large enough, in order to bring the electrom to its upper level. One may also notice that a smaller number of photons would be required for the $S_0 \rightarrow T_1$ transition. Nevertheless, the contribution of this transition to multiphoton absorption turns out to be negligible for NBOHC and Ge-ODC defects. Being theoretically forbidden because of symmetry reasons, its associated oscillation strength is, in fact, one million times smaller than the $S_0 \rightarrow S_1$ transition at room temperature \cite{girard}. Therefore, as a first approximation, we may only consider the presence of the $n$-photon absorption process. In this simplified model, the parameter $n$ represents an effective number of photons involved in the MPA process, since it includes all the possible contributions that we are not explicitly taking into account. We may also assume that the upconversion rate is small enough, so that the following relationship holds between the UL intensity ($I_{UL}$) and the $n$-th power of $P_p$ \cite{pollnau}:
\begin{equation}
    I_{UL} = \alpha P_p^n,
    \label{UL-intensity}
\end{equation}
where the parameter $\alpha$ includes the $n$-photon absorption cross-section, and the phase-matching condition between the laser pump and the luminescence radiation. In this sense, it worth noticing that the latter is isotropic, while the pump is mostly guided into the fiber core. Interestingly, the absorption bands corresponding to the NBOHCs and ODCs are quite close in energy. The first occurs at 258 nm, and the second at 241 nm \cite{girard}. Hence, UL generated by both defects is triggered by the same number of absorbed photons (specifically, $n=5$ for a pump wave at $\lambda$ = 1030 nm). To confirm this, in Fig.\ref{Spect_mod_n}c, we illustrated the dependence of $I_{UL}$ (calculated as the integral of the corresponding spectral peak as shown in Fig.\ref{Spect_mod_n}a) on $P_p$. Because of the overlap of the two ODC peaks at 400 and 460, we integrated them as if they were a single signal. Figure \ref{Spect_mod_n}c shows that the two fitting curves in the log-log plot are parallel lines with a slope equal to 5, thus confirming the prediction of the simple model of Eq.(\ref{UL-intensity}).

The estimation of the number of photons involved in the multi-photon absorption process allows us to properly model the intensity variation of UL along the propagation distance. The periodicity of the latter in GRIN fibers has already been pointed out as a result of the spatial-self imaging effect \cite{krupa,hansson,karlsson}. Nevertheless, the amplitude of the UL intensity variation along the propagation distance in a GRIN fiber could not be quantitatively reproduced. 

\begin{figure}[htbp]
  \centering
  \includegraphics[width=8.6cm]{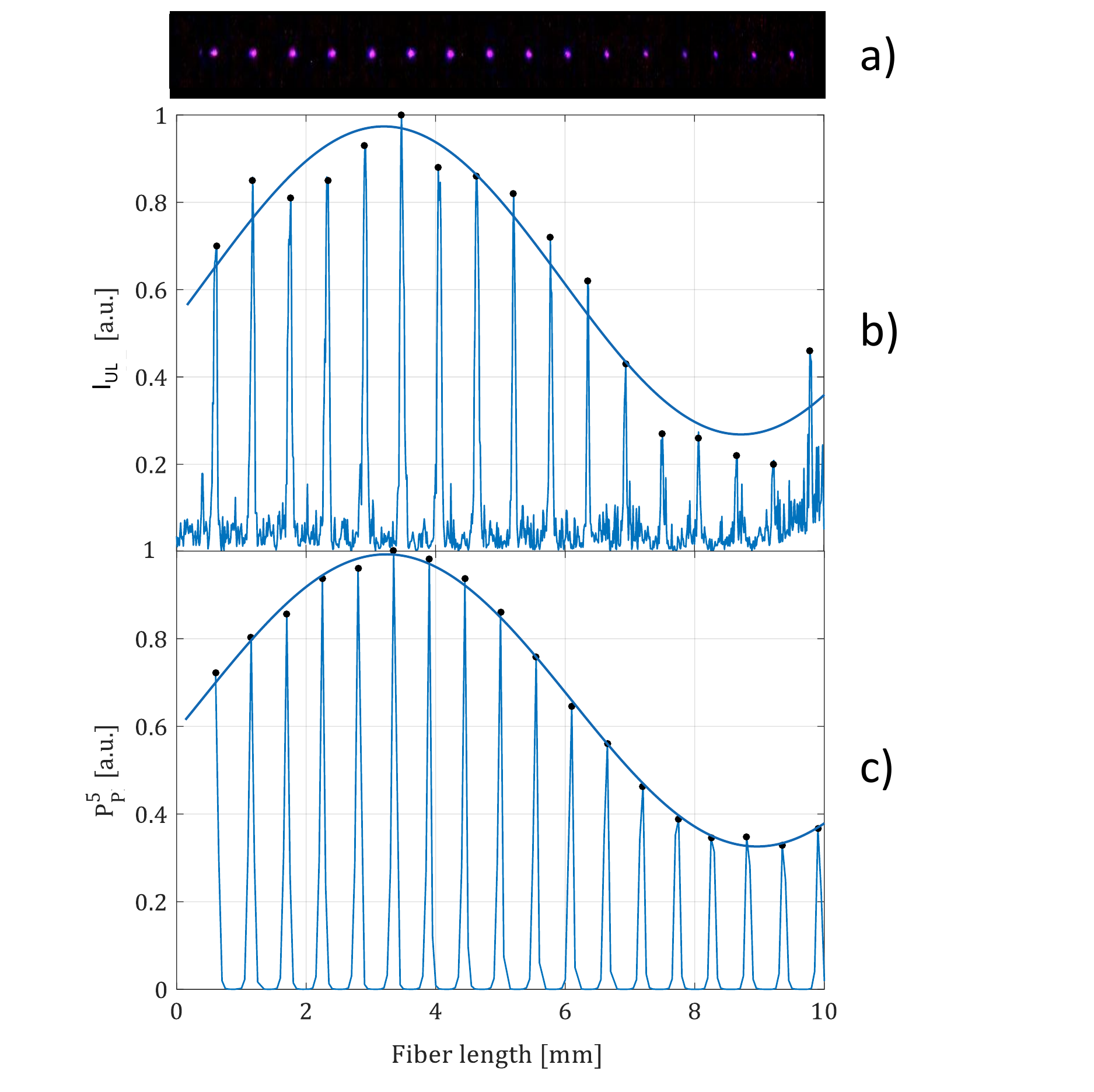}
\caption{(a) Digital microscope picture of light scattered from the GRIN fiber, showing periodic UL emission over the first centimeter. (b) 1D plot of the corresponding integrated normalized scattered intensity vs. propagation distance. The solid line is the envelope trend of the UL peaks (black dots) along with the fiber. (c) 1D plot of simulated fifth power of light intensity and its envelope.}
  \label{Selfimaging}
\end{figure}

In Figures \ref{Selfimaging}a,b we report a microscope image of the UL scattered over the first centimeter of GRIN fiber, and its corresponding intensity variation with distance. This was obtained by a normalization of the image intensity for each pixel of propagation. The presence of slow oscillation of the amplitude of the UL intensity peaks, whose first maximum occurs at about 3.5 mm, is highlighted by the solid line, which is drawn as a guide to the eye. Here the fiber length is limited to 10 mm: after that the metallic fiber holder introduces additional light scattering, which masks quantitative studies of the UL intensity. However, by employing longer samples, one notices a rapid quenching of visible light emission after the first few centimeters of fiber. To further check the relation (\ref{UL-intensity}) between UL and MPA, we performed a numerical simulation of spatiotemporal beam propagation in the fiber. We used a scalar 3D+1 GNLS equation including second, third, and fourth-order dispersion, Kerr, and Raman nonlinearities, as described in Ref.\cite{zitelli}. Random phase noise was added to the input field, in order to describe the generation of intensity speckles, and seed the generation of supercontinuum and dispersive waves. Random changes in the fiber core diameter were also considered, although their effect is negligible here because of the short fiber length. In Fig.\ref{Selfimaging}c we show the simulated fifth power of the light intensity peaks, along with their envelope (solid curve). As it can be seen by comparing  Fig.\ref{Selfimaging}b and Fig.\ref{Selfimaging}c, there is an excellent agreement between experimental and theoretical results.

\begin{figure}[htbp]
  \centering
  \includegraphics[width=9.4cm]{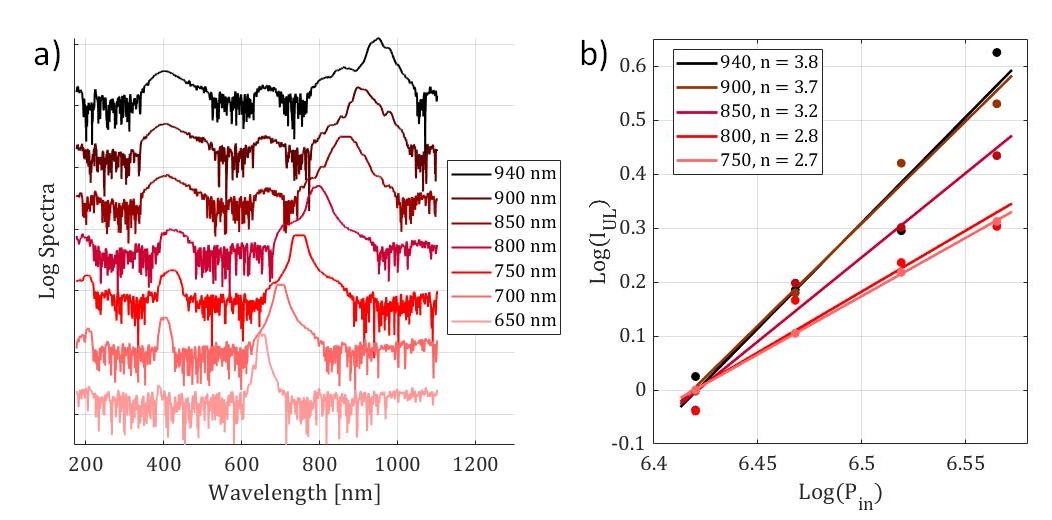}
\caption{(a) UL spectra in log scale, when the pump wavelength varies between 650 and 940 nm. (b) Loglog plot of UL intensity vs. pump wavelength. Experimental data (circles) are fitted by solid lines, whose slope $n$ appears in the legend.}
  \label{Comp_P_const}
\end{figure}

In order to further assess the validity of our model, we have varied the pump wavelength by using the OPA. In this way, the number of photons involved in the MPA process is also varied. We tuned the laser wavelength from 650 to 940 nm. In Fig.\ref{Comp_P_const}a we illustrate the variation of the UL spectrum with pump laser wavelength, when the input peak power is kept constant at 2.2 MW. The Ge-ODC spectral peaks at 400 and 460 nm are present and clearly visible for pump wavelengths above 750 nm. In Fig.\ref{Comp_P_const}b we plot the Ge-ODC UL intensity as a function of the input peak power. We avoid referring to the NBOHC luminescence at 650 nm, since it is partially covered by the spectral broadening of the source. Again, the linear trend in the log-log plots of Fig.\ref{Comp_P_const}b allows us to estimate the average number of photons that are involved in the absorption, as reported in the legend. As can be seen, the average number of photons is reduced with respect to the previous case: here $n$ ranges between 3.8 and 2.7 when the pump is swept from 940 to 750 nm. These values are in agreement with predictions of our model. Indeed, we expected the number of photons to decrease down to $n=3$ for $\lambda = 750$ nm, in order to match with the Ge-ODC absorption bands. Again, the discrepancy between theoretical and experimental values is rather small, in spite of our approximation that absorption is due to a single multiphoton process. The monotonic decrease of $n$ as the laser wavelength is reduced clearly indicates the MPA nature of the observed UL.

\begin{figure}[htbp]
  \centering
  \includegraphics[width=9cm]{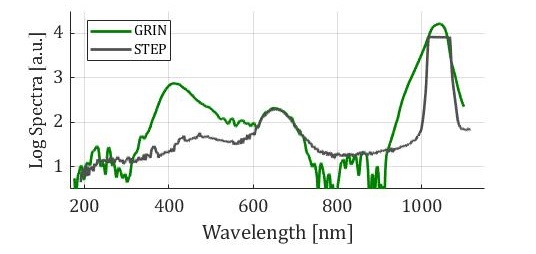}
\caption{Comparison of UL side-scattered spectra from either GRIN or step-index fibers, when the input peak power at 1030 nm is 2.2 MW. Curves plotted in a log scale are normalized to the same intensity value at 650 nm.}
  \label{Comp_STEP_GRIN}
\end{figure}
We also compared UL emission of GRIN and step-index MMFs, see Fig.\ref{setup}b. A typical side-scattered spectrum obtained from a step-index fiber is reported as a black solid curve in Fig.\ref{Comp_STEP_GRIN}. For the ease of comparison, we also display as a green solid curve the UL spectrum obtained at the same powerfrom a GRIN fiber, as in Fig.\ref{Spect_mod_n}a. Both spectra are normalized to the same intensity value at 650 nm. In this way, we can appreciate that the contribution of NBOHCs is much lower in a step-index fiber, since its spectral component at 1030 nm is higher (here its peak value is cut by the saturation of the spectrometer) than that from the GRIN fiber. We can attribute this to the absence of spatial self-imaging, which increases the number of emitting spots and, as a consequence, the total contribution of NBOHC defects to scattered light. As far as the ODCs are concerned, the UL signal is much less intense in a step-index fiber than in a GRIN fiber. This can be attributed to the lack of extrinsic Ge atoms. Nevertheless, one can still easily recognize the presence of two peaks, which are slightly shifted with respect their respective positions for the GRIN fiber. We believe that these peaks are due to intrinsic Ge impurities that are present in small amounts even in undoped commercial optical fibers \cite{messina-2008}. However, it worth noticing that Si-ODCs and Ge-ODCs have almost overlapped emission spectra and comparable oscillation strength \cite{girard}. Therefore, UL between 400 and 500 nm in the step-index fiber may also have a contribution from Si-associated defects.

In conclusion, we reported the observation of MPA excited UL in commercial multimode optical fibers. Based on our measurements, we estimate that MPA involving an average number of up to $n=5$ photons is at the origin of the observed UL, in agreement with the expected position of the absorption bands. We could also directly relate the observed variation of side-scattered UL light intensity in GRIN fibers with simulated variation of MPA strength along the fiber. In perspective, UL can be exploited for extending the bandwidth of supercontinuum generation into UV frequencies. Moreover, the appearance of UL provides a useful tool to directly monitor the presence of nonlinear losses, at peak powers close to the damage threshold of optical fibers. 
This paves the way for future investigations of peak power limitations in spatiotemporal mode-locking with multimode fiber lasers.

\section*{Acknowledgments}
We acknowledge support from the European Research Council (ERC) under the European Union’s Horizon 2020 research and innovation program (grant No. 740355). The authors declare no conflicts of interest.

\bibliographystyle{unsrt}

\end{document}